\documentclass[review]{elsarticle}

\usepackage{tikz}
\usepackage[framemethod=TikZ]{mdframed}
\usepackage[RPvoltages]{circuitikz}
\usetikzlibrary{shapes,spy}
\usepackage{amsmath}
\usepackage{amssymb}
\usepackage{fancyhdr}
\usepackage{mathtools}
\usepackage{dsfont}
\usepackage{braket}
\usepackage[a4paper, left=3cm,right=3cm, top=4cm]{geometry} 
\usepackage[OT2,T1]{fontenc}
\usepackage{textcomp}
\usepackage{hyperref}
\usepackage{pdfcomment}
\usepackage{longtable}
\usepackage{booktabs}
\usepackage{dcolumn}
\usepackage{empheq}
\usepackage[many]{tcolorbox}
\usepackage[english]{babel}
\usepackage{csquotes} 

\DeclareSymbolFont{cyrletters}{OT2}{wncyr}{m}{n}
\DeclareMathSymbol{\Sha}{\mathalpha}{cyrletters}{"58}

\newcommand{\textcite}[1]{\cite{#1}}

\mdfdefinestyle{EndFrame}{%
    linecolor=black,
    outerlinewidth=0.35pt,
    roundcorner=5pt,
    innertopmargin=0,
    innerbottommargin=\baselineskip,
    innerrightmargin=6.5pt,
    innerleftmargin=6.5pt
    }

\AtBeginDocument{\renewcommand{\natexlab}[1]{#1}}

\begin{document}

\title{On Boundary Conditions in the sub-mesh interaction of the Particle-Particle-Particle-Mesh Algorithm}
\author[1]{Tim Wyssling}
\author[2]{Andreas Adelmann\corref{cor1}}
\ead{andreas.adelmann@psi.ch}
\cortext[cor1]{Corresponding author}
\address[1]{Paul Scherrer Institut, Forschungsstrasse 111, 5232 Villigen, Switzerland.}
\address[2]{Paul Scherrer Institut, Forschungsstrasse 111, 5232 Villigen, Switzerland.}
\begin{abstract}
The Particle-Particle-Particle-Mesh algorithm elegantly extends the standard Particle-In-Cell scheme by direct summation of interaction that happens over distances below or around mesh size. Generally, this allows for a more accurate description of Coulomb interactions  and improves precision in the prediction of key observables. Nevertheless, most implementations neglect electrostatic boundary conditions for the short-ranged interaction that are directly summed. In this paper a variational description of the Particle-Particle-Particle-Mesh algorithm will be developed for the first time and subsequently used to derive temporally and spatially discrete equations of motion. We show that the error committed by neglecting boundary conditions on the short scale is directly tied to the discretization error induced by the computational grid.
\end{abstract}


\maketitle 

\section{Introduction}

Coulomb interactions, also known as space-charge effects, have a tangible influence on charged particle  dynamics, which manifests itself in effects such as emittance growth, particle loss or halo formation. The emittance is a measure of the volume of particles in phase space and with halo we refer to particles which are spatially several sigmas away from the center of charge. Especially for low-emittance particle sources it is crucial to have an accurate and precise model for space-charge effects in order to make accurate predictions for potential applications.

 One of the challenge, in resolving the dynamics of charged $N$-body systems, is the efficient and accurate computation of the Coulomb interaction. The computational cost of summing up the Coulomb interaction pair-wise in an $N$-body system scales with $\mathcal{O}(N^2)$. A large number of beam particles inevitably necessitates a trade-off between accuracy and computational speed, which has brought forth a multitude of techniques and schemes.

The Particle-In-Cell (PIC) scheme is a well-known method to include space-charge effects self-consistently. It relies on computational grids to discretize the charge density resulting from  a finite number of particles. The electric potential is obtained by solving Poisson's equation on the grid for each time step, from which the forces can be obtained via finite differences. The PIC scheme scales, at best, linearly with $\mathcal{O}(N)$ and is the method of choice for many simulations \cite{hockney_computer_1988}. The lower computational cost comes at the price of errors in the forces that are a consequence of the interpolation to and from the computational grid. The spatial resolution of the potential is ultimately restricted by the smallest grid length, which is problematic when the particles are closer and leads to numerical errors \cite{hejlesen_non-singular_2019,nevins_discrete_2005,qiang_symplectic_2017}. 

The Particle-Particle-Particle-Mesh (P3M) algorithm \cite{hockney_computer_1988} overcomes the discretization error in the force by computing the potential for close particles through direct summation and through PIC for particles that are farther apart. In other words the interaction that happens on a scale smaller than the mesh size is reintroduced by direct summation. Conceptually, the P3M scheme can be understood as an Ewald summation on a grid that uses discrete Fourier transformation for efficient evaluation. 

\tcbset{highlight math style={enhanced,
		colframe=red,colback=white,arc=4pt,boxrule=2pt,
}}

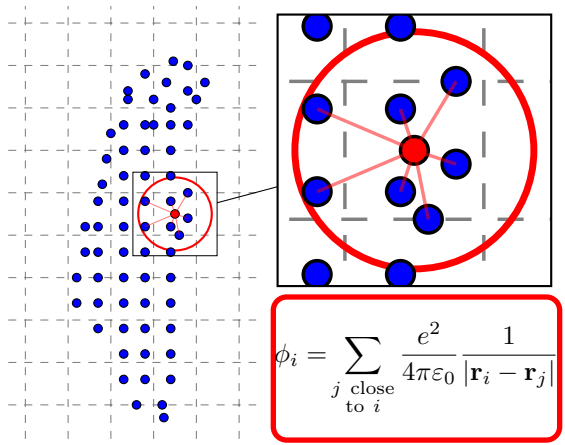
\begin{figure}[htpb]
\centering
\begin{tikzpicture}[spy using outlines={rectangle,black,magnification=3.25,size=3.6cm, connect spies}]
    \begin{scope}[xscale=0.9, yscale=0.9]
    
    \begin{scope}[xscale=1.25, yscale=1.25]
	\draw[step=.5cm, color=gray, style=dashed](1.3,-1.4) grid (4.2,3.7);
	\draw [red, thick] (3.25,1.25) circle [radius=0.43];
	\foreach \a in {2.65, 2.9, 3.2}
	\foreach \b in {-0.7,-0.4,...,2.5} {
		\draw [fill=blue] (\a,\b) circle [radius=0.05];
	}
	\foreach \a in {2.35}
	\foreach \b in {-0.1,0.2,...,1.7}
	\draw [fill=blue] (\a,\b) circle [radius=0.05];
	
	\foreach \a in {(3,2.3),(3.2,2.5),(3.0,2.6),(3.15,2.8),(2.9,2.9),(3.23,3.05), (3.4, 3.0), (3.6,2.8), (3.5, 2.6), (2.7, 2.7), (2.7,2.6), (2.45,1.9), (2.4,1.6), (2.5,2.2),(2.1, 0.2), (2.1, 0.5), (2.2, 0.8), (2.8, -1.0), (3.1, -1.0),(3.12, -1.15), (2.2, 1.1), (3.4,2.7), (3.4, 2.3), (3.4, 1.5), (3.4,1.2), (3.3,1.0)}
	\draw [fill=blue] \a circle [radius=0.05];
	\draw [fill=red] (3.25,1.25) circle [radius=0.05];
	
	\draw [-, color=red, opacity=0.5] (3.25,1.25) to (3.3,1.0);
	\draw [-, color=red, opacity=0.5] (3.25,1.25) to (3.4,1.2);
	\draw [-, color=red, opacity=0.5] (3.25,1.25) to (3.4, 1.5);
	\draw [-, color=red, opacity=0.5] (3.25,1.25) to (3.2, 1.1);
	\draw [-, color=red, opacity=0.5] (3.25,1.25) to (3.2, 1.4);
	\draw [-, color=red, opacity=0.5] (3.25,1.25) to (2.9, 1.1);
	\draw [-, color=red, opacity=0.5] (3.25,1.25) to (2.9, 1.4);
	
    \end{scope}
    
    \spy on (3.65625,1.40625) in node [right] at (5.00625,2.25);
    
    \node [align=left] at (7.35,-0.5) {\begin{minipage}{3.4cm} \begin{empheq}[box=\tcbhighmath]{align}
	\phi_i = &\sum_{\substack{j\ \text{close} \\ \text{to}\ i}} \frac{e^2}{4\pi\varepsilon_0} \frac{1}{|\mathbf{r}_i - \mathbf{r}_j|} \notag 
	\end{empheq} \end{minipage}};

    \end{scope}
	\end{tikzpicture}
\caption{The interaction with close neighbouring particles is summed directly. The interaction with farther away particles is computed one the computational grid via PIC.}
\label{fig:p3m_concept}
\end{figure}

In most models the short-ranged part is computed by directly summing the interaction using Coulomb's law, which is then added to the particle equation of motion. Formally, the short-ranged potential, as seen by the $i$-th particle, reads,

\begin{equation}
\phi_i = \sum_{\substack{j\ \text{close} \\ \text{to}\ i}} \frac{q^2}{4\pi\varepsilon_0} \frac{1}{|\mathbf{r}_i - \mathbf{r}_j|} ,
\label{eq:summation_short}
\end{equation}

where it was used that the Coulomb potential $\phi$ in between two particles with distance $r$ and charge $q$ is given as $\phi(r) = \frac{q^2}{4\pi\varepsilon_0} \frac{1}{r}$. This formula for the potential comes from solving Poisson's equation, where the term $\frac{1}{r}$ is Green's function. However, it is only viable in the absence of electrostatic boundary conditions, which would require a different Green's function. In other words, by direct summation using this formula, the influence of electrostatic boundary conditions is neglected for close particle interaction, especially in the presence of conductors, such as beam tubes or cathodes. Up to now, it is unclear to which degree this assumption can be made.

This paper serves the purpose of investigating this error.

\section{Equations of Motion}

In order to quantify the error we require a solid theoretical formulation of the P3M procedure, that properly accounts for spatial and temporal discretization. In the following section we will introduce a Lagrangian description, which allows to derive fully discrete Poisson's equation, including the short and long range potentials. This will allow to properly introduce boundary conditions into eq (\ref{eq:summation_short}) and examine their impact in detail.

\subsection{Lagrangian Beam Dynamics}

The Lagrangian approach has established itself as the method of choice to derive equations of motion in nearly all fields of physics. In beam dynamics in particular it is routinely employed to derive phase space maps for beam line elements. The main advantage next to assured physical validity is the symplectic nature of the equations derived \cite{qiang_symplectic_2017,qiang_symplectic_2018}, which is beneficially related to conservation laws of e.g. energy and momentum. A widely known Lagrangian description stems from \textcite{low_lagrangian_1958}, whose derivation (but not result) was later corrected by \textcite{galloway_lagrangian_1971}. It is not the only possibility for a Lagrangian \cite{hirvijoki_action_2016} but reproduces the physical equations of motions nevertheless, as shown by \textcite{evstatiev_variational_2013}, whose notation will be used in the following. 

The aforementioned Lagrangians are all conceptually related to Vlasov's equation in the sense that they assume implicitly a small value (below 0.1) of the plasma coupling factor $\Gamma = E_C/(k_BT)$, where $E_C$ is the electrostatic energy. Our main goal is to derive a Lagrangian that incorporates the summation procedure of the P3M algorithm and thus circumvents this shortcoming.

To establish the formalism we denote the 6-dimensional phase space distribution as $f(\mathbf{x},\mathbf{v},t)=f(\mathbf{x}(t),\mathbf{v}(t),t)$. The quantities $\mathbf{x}$, $\mathbf{v}$ are themselves solutions to the equations of motion and only dependent on time $t$ and on initial conditions $\mathbf{x}_0$, $\mathbf{v}_0$. If Liouville's theorem holds, we can express the distribution at time $t$ through an initial distribution $f_0$ as $f(\mathbf{x}(t),\mathbf{v}(t),t)=f_0(\mathbf{x}_0,\mathbf{v}_0)$, i.e. there is no time dependence. With this in mind, Low's Lagrangian reads as follows,
\begin{equation}
\begin{aligned}
\mathcal{L}(\mathbf{x}, \mathbf{v},\phi) &= \int d^3\mathbf{x}_0\int d^3\mathbf{v}_0\ f_0(\mathbf{x}_0,\mathbf{v}_0)\left[\frac{1}{2}m\mathbf{v}^2(\mathbf{x}_0,\mathbf{v}_0) - q\phi(\mathbf{x}(\mathbf{x}_0,\mathbf{v}_0))\right] + \frac{\varepsilon_0}{2}\int d^3\mathbf{x} (\nabla\phi)^2(\mathbf{x}(\mathbf{x}_0,\mathbf{v}_0)) \\
&= \mathcal{L}_p + \mathcal{L}_\text{int} + \mathcal{L}_f,
\end{aligned}
\label{eq:lowlagrangian}
\end{equation}
where the first and last term represent particle and field energy respectively and q represents the charge. The interaction term can be simplified considering that after integrating $f$ over velocity we are left with the spatial charge distribution,
\begin{align*}
    \int d^3\mathbf{x}\ \rho(\mathbf{x}) &= \int d^3\mathbf{x}_0\ \int d^3\mathbf{v}_0\ qf_0(\mathbf{x}_0, \mathbf{v}_0) \\ &= \int d^3\mathbf{x}\ \int d^3\mathbf{v}\ qf(\mathbf{x},\mathbf{v}),
\end{align*}
where it was used, that,
\begin{equation}
	f_0(\mathbf{x}_0,\mathbf{v}_0)d^3\mathbf{x}_0d^3\mathbf{v}_0 = f(\mathbf{x},\mathbf{v})d^3\mathbf{x}d^3\mathbf{v},
	\label{eq:part_num_cons}
\end{equation}
which corresponds to particle number conservation for a phase space element \cite{galloway_lagrangian_1971}, which is a consequence of Liouville's theorem.

The equations of motion can be obtained by variation after $\mathbf{x}$ and $\phi(\mathbf{x})$ respectively,
\begin{align*}
m\ddot{\mathbf{x}} &= -q\nabla_x\phi(\mathbf{x}) ,\\
\nabla^2_x\phi(\mathbf{x}) &= \frac{1}{\varepsilon_0}\int d^3\mathbf{v}\ f(\mathbf{x},\mathbf{v}),\ q =:\frac{\rho(\mathbf{x})}{\varepsilon_0}.
\end{align*}
These equations describe the self-consistent electromagnetic dynamics that underpin space charge effects.

The solution to Poisson's equation can be obtained by convolution with Green's function $G$,
\begin{equation}
\phi(\mathbf{x}) = \frac{1}{\varepsilon_0} [\rho \otimes G] (\mathbf{x}),
\label{eq:greens_function_for_poisson}
\end{equation}
where $\otimes$ denotes the convolution,
$$\frac{1}{\varepsilon_0} [ \rho \otimes G] (\mathbf{x}) = \frac{1}{\varepsilon_0} \int d^3\mathbf{y}\ \rho(\mathbf{y}) G(\mathbf{x}-\mathbf{y}).$$ If there are boundary conditions they are reflected in $G$. For the case of an unbound domain, Green's function assumes the following well-known form,
\begin{equation}G(\mathbf{x},\mathbf{y}) = \frac{1}{|\mathbf{x}-\mathbf{y}|}.
\label{eq:free_gf}
\end{equation}

\subsection{Ewald Splitting}
In the following we will introduce a way to mathematically describe the P3M scheme without the need to refer to summing techniques. This will be used to motivate an Ansatz for a corresponding Lagrangian.

As illustrated in figure \ref{fig:p3m_concept}, P3M distinguishes between short- and long-ranged potential contributions, where the former is summed directly and latter is computed via PIC. This splitting is conceptually rooted in Ewald's summation technique and can be represented by splitting the interaction potential of Coulomb's law,
$$\frac{1}{r} = \frac{1-f(r)}{r} + \frac{f(r)}{r}.$$
In the context of electromagnetic interaction, one usually chooses $f$ as the error function,
$$f(r) = \text{Erf}(\alpha r) = \frac{1}{2\sqrt{\pi}}\int_{\alpha r}^\infty dt\ e^{-t^2},$$
which gives,
\begin{equation}
\begin{aligned}
\frac{1}{|\mathbf{r}_i-\mathbf{r}_j|} = & \frac{\text{Erfc}(\alpha|\mathbf{r}_i-\mathbf{r}_j|)}{|\mathbf{r}_i-\mathbf{r}_j|} +\frac{\text{Erf}(\alpha|\mathbf{r}_i-\mathbf{r}_j|)}{|\mathbf{r}_i-\mathbf{r}_j|}.
\end{aligned}
\label{eq:ewald_split}
\end{equation}
by using $\text{Erfc}(x) = 1 - \text{Erf}(x)$.
One can see that the first term of eq (\ref{eq:ewald_split}) is short-ranged and singular, whereas the second term is long-ranged and regular. The Ewald parameter $\alpha$ serves to control the size of the Ewald sphere, outside of which the short-ranged term vanishes. Since the long-ranged interaction is regularized to the size of the Ewald sphere it can be solved using a computational mesh without the errors that the singularity would normally cause in a discrete computation. The singularity in the short-ranged interaction can be handled by omitting the self-energy in the summation.

Albeit very straightforward, this way of splitting becomes non-trivial in the presence of boundary conditions, where the interaction potential is not given by $\frac{1}{r}$. In other words the splitting procedure also means that Green's function is no longer a fundamental solution to Laplace's operator as appearing in Poisson's equation. Since we are primarily interested in the role boundary conditions play, we will entertain an alternative approach. Instead of splitting the interaction potential (or more precisely: Green's function), one can equivalently split the charge distribution \cite{deserno_how_1998,deserno_how_1998_1}. Through the subtraction and subsequent superposition
of a screened charge distribution, we obtain the same result. Let $\gamma$ be the screening function, we can write the charge distribution $\rho$ as follows,
\begin{equation}
\begin{aligned}
    \rho & = \rho - \rho\otimes\gamma + \rho\otimes\gamma , \\
    \rho^ S & = \rho - \rho\otimes\gamma , \\
    \rho^L &= \rho\otimes\gamma ,
\end{aligned}
\label{eq:charge_splitting}
\end{equation}
where the short-hand notations $\rho^S$, $\rho^L$ were introduced to succinctly denote screened charge and the screening charge, respectively. 
In terms of the potential this is equivalent, because it is given as the convolution between Green's function and charge distribution as in eq (\ref{eq:greens_function_for_poisson}),
\begin{align*}
\phi &= G \otimes \rho \\
&= G \otimes (\rho - \rho\otimes\gamma + \rho\otimes\gamma) \\
&= G\otimes\rho - G\otimes\gamma\otimes\rho + G\otimes\gamma\otimes\rho \\
&= \left(G- G\otimes\gamma + G\otimes\gamma\right)\otimes\rho.
\end{align*}
If $\gamma$ is chosen in the form of a Gaussian \cite{deserno_how_1998}, 
$$\gamma = \frac{\alpha^3}{\sqrt{\pi^3}}e^{-\alpha^2r^2},$$
and $G$ is taken as in eq (\ref{eq:free_gf}) we obtain the same splitting with the error function as in eq (\ref{eq:ewald_split}).

Since Poisson's equation is linear, we may write the potential as a sum of two potentials originating from two charge distributions,
\begin{equation}
    \begin{aligned}
        \phi &= \phi_{\rho-\rho\otimes\gamma} + \phi_{\rho\otimes\gamma} , \\
	    \phi^S &= \phi_{\rho-\rho\otimes\gamma}, \\
	    \phi^L &= \phi_{\rho\otimes\gamma} ,
    \end{aligned}
	\label{eq:pot_split}
\end{equation}
where the short-hand notations $\phi^S$, $\phi^L$ were introduced to succinctly denote short-ranged and long-ranged potentials.

Since the entire splitting is contained in the charge distribution, Green's function does not require modifications with the error function as in eq (\ref{eq:ewald_split}). Instead Green's function is still a fundamental solution to the Laplacian, and one can use whatever form is appropriate for the boundary conditions at hand. 

\subsection{Lagrangian Formulation of P3M}
Starting again from eq (\ref{eq:lowlagrangian}), Low's Lagrangian, we make the following Ansatz using eq (\ref{eq:pot_split}):
\begin{equation}
\begin{aligned}
& \mathcal{L}(\mathbf{x},\dot{\mathbf{x}},\phi^S, \phi^L) \\ &= \int d^3\mathbf{x}_0d^3\mathbf{v}_0 f(\mathbf{x}_0,\mathbf{v}_0) \ \frac{m}{2}\dot{\mathbf{x}}^2(\mathbf{x}_0,\mathbf{v}_0, t) \\
&\quad \int d^3\mathbf{x} \biggl[-\rho^S(\mathbf{x})\phi^S(\mathbf{x})-\rho^L(\mathbf{x})\phi^L(\mathbf{x})\biggr] \\ &\quad + \frac{\varepsilon_0}{2}\int d^3\mathbf{x}\ (\nabla\phi^S)^2 + \frac{\varepsilon_0}{2}\int d^3\mathbf{x}\ (\nabla\phi^L)^2.
\end{aligned}
\label{eq:Lagrangian_Ansatz}
\end{equation}
The main idea of this Ansatz is to treat both potentials as two distinct functions, which is required since P3M computes each in a different way. At first glance this might seem inconsistent as we cannot obtain eq (\ref{eq:Lagrangian_Ansatz}) simply by inserting the splitting (\ref{eq:pot_split}) into Low's Lagrangian (\ref{eq:lowlagrangian}), due to the quadratic field energy terms. However, this Lagrangian will also yield two distinct Poisson's equations in the form of the Euler-Lagrange equations for $\phi^S$ and $\phi^L$. In order to motivate the Ansatz, we can anticipate that the solutions to these equations, i.e. the potentials, will fulfil eq (\ref{eq:pot_split}) again.

\subsection{Spatial and Temporal Discretization}
Equation (\ref{eq:Lagrangian_Ansatz}) contains the phase space probability density function and the electric potentials, which are both treated as continuous fields. However, the P3M algorithm prescribes that the long-ranged interaction $\phi^L$ is obtained via PIC on a mesh, which must be accommodated by introducing a spatial discretisation. In addition both potentials are computed at discrete times. In the following it will be shown how the Lagrangian can be discretized to yield practical equations of motion. 

\subsubsection{Macro-particles}

In case of a continuous distribution it may be necessary to decompose the distribution function $f$ into $N$ macro-particles $f_\alpha$, where $\alpha=1 ... N$. The $\alpha$-th macro-particle has a finite spatial shape $S_x$, a finite velocity shape $S_v$, a weight $\omega_\alpha$ and is located at the position $\boldsymbol{\xi}_\alpha$,
\begin{equation}
\begin{aligned}
f(\mathbf{x},\mathbf{v}) &= \sum_\alpha f_\alpha(x,v,t) \\
&= \sum_\alpha \omega_\alpha S_x(\mathbf{x}-\boldsymbol{\xi}_\alpha)S_v(\dot{\mathbf{x}}-\dot{\boldsymbol{\xi}}_\alpha) .
\end{aligned}
\label{eq:macroparticles}
\end{equation}

Usually, the $\omega_\alpha$ are chosen as the amount of physical particles per macroparticle, $S_v$ is a $\delta$-function and $S_x$ is a B-spline function of order 0 or 1. In the limit of small particle numbers this is not necessary and we can replace $S_x$, $S_v$ with $\delta$ functions accordingly. There are only a few conditions that the shape functions must fulfil. Additionally to being symmetric and compact, they should be normalized to 1,
$$\int_{-\infty}^\infty S_x(\mathbf{x}-\boldsymbol{\xi})d^3\mathbf{x} = 1 = \int_{-\infty}^\infty S_v(\mathbf{v}-\dot{\boldsymbol{\xi}})d^3\mathbf{v}.$$

The charge distribution $\rho(\mathbf{x})$ is now given as the charge contribution of every particle $\boldsymbol{\xi}_\alpha$ to the point $\mathbf{x}$. Formally, this reads,

\begin{equation}
    \begin{aligned}
        \rho(\mathbf{x}) &= \sum_\alpha \omega_\alpha S_x(\mathbf{x}-\boldsymbol{\xi}_\alpha) ,\\
        \rho^S(\mathbf{x}) &= \sum_\alpha \omega_\alpha \bigl[S_x - S_x\otimes\gamma\bigr](\mathbf{x}-\boldsymbol{\xi}_\alpha) ,\\
        \rho^L(\mathbf{x}) &= \sum_\alpha \omega_\alpha \bigl[S_x\otimes\gamma\bigr](\mathbf{x}-\boldsymbol{\xi}_\alpha).
    \end{aligned}
    \label{eq:macropart_charge_dists}
\end{equation}

\subsubsection{Spatial Discretization}

For the introduction of a computational spatial grid, let us assume an euclidean coordinate space $[0,L_x]\times[0,L_y]\times[0,L_z]$ that is discretized as $L_x=N_xh_x$, $L_y=N_yh_y$, $L_z=N_zh_z$, where $N$ and $h$ give the amount of steps and the stepsize in the respective direction. The coordinates can then be denoted as $x_k = k\cdot h_x$, $y_l = l\cdot h_y$, $z_m = m\cdot h_z$ and likewise the electromagnetic potential $\phi(x,y,z)$ is given only at a discrete set of mesh-points $\phi_{klm} = \phi(x_k,y_l,z_m) = \phi(k\cdot h_x,l\cdot h_y,m\cdot h_z)$. To reconcile this with the continuous Lagrangian description (\ref{eq:Lagrangian_Ansatz}), we can use functions $I(x)$ to interpolate in between grid points \cite{evstatiev_variational_2013},

\begin{align*}
    \phi^L(\mathbf{x}) = \sum_{klm}(\phi^L)_{klm} & I_k\left(\frac{x_k-x}{h_x}\right)I_l\left(\frac{y_l-y}{h_y}\right) \\ &\cdot I_m\left(\frac{z_m-z}{h_z}\right).    
\end{align*}

In the sequel, multi-index notation $i=(k,l,m)$ is used, to write more succinctly as,
\begin{equation}
\phi^L(\mathbf{x}) = \sum_{i}(\phi^L)_{i}I_i\left(\frac{\mathbf{x}_i-\mathbf{x}}{\Delta\mathbf{x}}\right).
\label{eq:grid_potential}
\end{equation}

The simplest choice for $I_i$ are B-splines of 0th order, which essentially corresponds to assume a constant charge distribution over a computational cell. This can be seen from the definition,
\begin{equation*}
\begin{aligned}
    I_i\left(\frac{\mathbf{x}_i-\mathbf{x}}{\Delta\mathbf{x}}\right) &= b_0\left(\frac{\mathbf{x}_i-\mathbf{x}}{\Delta\mathbf{x}}\right) \\  &= \begin{cases}
    1 \quad \text{if}\quad \left(\frac{\mathbf{x}_i-\mathbf{x}}{\Delta\mathbf{x}}\right) < \frac{1}{2} ,\\
    0 \quad \text{otherwise}.
\end{cases}
\end{aligned}
\end{equation*}

Additionally, the derivatives in the field energy term of (\ref{eq:Lagrangian_Ansatz}) must be approximated using finite differences. An x-directed finite difference operator of second order $D_x$ acting on a discrete 3-dim potential $\phi_{jlm}$ can be defined as \cite{berger_use_1958},
\begin{equation*}
	\begin{aligned}
	(D_{x} \phi)_{klm} &= \sum_j (D_{x})_{kj}\phi_{jlm} \\ & =  \phi_{(k+1)lm}+\phi_{(k-1)lm}-2\phi_{klm} \\
	(D_x)_{ji} &= \delta_{j(i+1)}+\delta_{j(i-1)}-2\delta_{ji}.
	\end{aligned}
	\label{eq:1d_ddiff}
\end{equation*}
Through the way the mesh was defined, it was already implicitly assumed, that the coordinates are separable.\ Thus the multi-dimensional discrete Laplacian can be constructed directly through a Kronecker-sum of discrete one-dimensional Laplacians,
\begin{equation}
    D_{xyz} = D_{x}\otimes\mathds{1}_{y}\otimes\mathds{1}_{z} + \mathds{1}_{x}\otimes D_{y}\otimes\mathds{1}_{z} + \mathds{1}_{x}\otimes\mathds{1}_{y}\otimes D_{z},
    \label{eq:3dim_second_order_fin_diff_tensor}
\end{equation}

where $\otimes$ denotes the tensor product. This assumption holds as long as the respective coordinate system is not curved. The matrix elements of this operator can analogously be defined as a sum of Kronecker-Deltas,
\begin{equation}
\begin{aligned}
    (D_{xyz})_{abc;def} = (D_x)_{ab}\delta_{cd}\delta_{ef} & +\delta_{ab}(D_y)_{cd}\delta_{ef} \\ & +\delta_{ab}\delta_{cd}(D_z)_{ef}.
\end{aligned}
\label{eq:3dim_second_order_fin_diff}
\end{equation}
Before discretization of the interaction term of (\ref{eq:Lagrangian_Ansatz}), one can use partial integration to achieve second order accuracy, which allows to directly use a second order finite differences scheme
\begin{equation}
\begin{aligned} \label{eq:discrete_field_term}
& \int d^3\mathbf{x}\ (\nabla\phi^L(\mathbf{x}))\cdot(\nabla\phi^L(\mathbf{x})) ,\\ &= -\int d^3\mathbf{x}\ \phi^L(\mathbf{x})\Delta\phi^L(\mathbf{x}) ,\\
&= -h_xh_yh_z\sum_{klm} (\phi^L)_{klm}(\Delta(\phi^L)(\mathbf{x}))_{klm} ,\\
&= -h_xh_yh_z\sum_{klm}\sum_{abc}(\phi^L)_{klm}(D_{xyz})_{klm;abc}(\phi^L)_{abc} ,\\
&= -h_xh_yh_z\sum_{i} \sum_{j} (\phi^L)_{i} (D_{xyz})_{i;j}(\phi^L)_{j},
\end{aligned}
\end{equation}
where again $i$ and $j$ are multi-indices.

Alongside the field the charge distribution also became discrete. This can be seen by looking at the long-range interaction term of (\ref{eq:Lagrangian_Ansatz}),
\begin{align*}
\mathcal{L}_\text{int} &= \sum_i (\phi^L)_i \int d^3\mathbf{x}\ I_i\left(\frac{\mathbf{x}_i-\mathbf{x}}{\Delta\mathbf{x}}\right) \rho^L(\mathbf{x})  ,\\
&= \sum_i (\phi^L)_i (\rho^L)_i  ,\\
(\rho^L)_i &= \int d^3\mathbf{x}\ I_i\left(\frac{\mathbf{x}_i-\mathbf{x}}{\Delta\mathbf{x}}\right) \rho^L(\mathbf{x}).
\end{align*}
Using (\ref{eq:macropart_charge_dists}) we can define the interpolation function,
\begin{equation}
    W^L(\mathbf{x}_i-\boldsymbol{\xi}_\alpha) = \int d^3\mathbf{x}\ \bigl[S_x\otimes\gamma\bigr](\mathbf{x}-\boldsymbol{\xi}_\alpha) I_i\left(\frac{\mathbf{x}_i-\mathbf{x}}{\Delta\mathbf{x}}\right),
\label{eq:WL}
\end{equation}
between particles in continuous space and the grid. If we choose $S_x=b_0$, $I_i = b_0$ we regain cloud-in-cell interpolation via the recursion relation for splines,
\begin{equation}
\begin{aligned}
    W\left(\frac{\mathbf{x}_i-\boldsymbol{\xi}_\alpha}{\Delta\mathbf{x}}\right) &= \int d^3 \mathbf{x}\ b_0\left(\frac{\mathbf{x}-\boldsymbol{\xi}_\alpha}{\Delta\mathbf{x}}\right)b_0\left(\frac{\mathbf{x}_i-\mathbf{x}}{\Delta\mathbf{x}}\right) \\ &= b_1\left(\frac{\mathbf{x}_i-\boldsymbol{\xi}_\alpha}{\Delta\mathbf{x}}\right) .
\end{aligned}
\label{eq:interpolation_splines_b0}
\end{equation}
For the sake of consistency and ease of notation we can also define a short-range ''interpolation'' function,
\begin{equation}
	W^S(\mathbf{x}-\boldsymbol{\xi}_\alpha) = \bigl[S_x-S_x\otimes\gamma\bigr]  (\mathbf{x}-\boldsymbol{\xi}_\alpha),
	\label{eq:WS}
\end{equation}
which does not actually contain interpolation, since the short-ranged potential is not computed on the grid.

\subsubsection{Temporal Discretization}
Every numerical scheme that solves a differential equation of motion presumes a discrete stepping in time. In the following, it will be shown how discrete time stepping can be woven into the Lagrangian by using variational integrators \cite{kraus_variational_2014,lew_overview_2003,rowley_variational_2002}. The basic idea is that the optimal trajectory - which minimizes the action as described by Hamilton's principle -  is approximated via a discrete set of patches $\mathcal{L}_d$. The action functional is approximated by a sum over M intervals with size $\Delta t$,
\begin{align*}
S &= \int dt\  \mathcal{L}(\boldsymbol{\xi}(t),\dot{\boldsymbol{\xi}}(t), \phi(\boldsymbol{\xi}(t),t)) \\
&\approx \sum_{n=0}^{M-1} \int_{t_n}^{t_{n+1}}dt\ \mathcal{L}(\boldsymbol{\xi}(t),\dot{\boldsymbol{\xi}}(t), \phi(\boldsymbol{\xi}(t),t)) \\
&\approx \sum_{n=0}^{M-1}  \mathcal{L}_d(\boldsymbol{\xi}^n,\boldsymbol{\xi}^{n+1}, \phi^n,\phi^{n+1},\Delta t),
\end{align*}
where,
\begin{align*}
\boldsymbol{\xi}^n &= \boldsymbol{\xi}(n\cdot \Delta t), \\
\phi^n &= \phi(\boldsymbol{\xi}(n\cdot \Delta t), n\cdot \Delta t)\ \text{ and} \\
\mathcal{L}_d(...,\Delta t) &= \int_{t_n}^{t_{n+1}}dt\  \mathcal{L}(\boldsymbol{\xi}(t),\dot{\boldsymbol{\xi}}(t), \phi(\boldsymbol{\xi}(t),t), t).
\end{align*}

For ease of notation we have omitted the macro-particle index $\alpha$ from $\mathbf{\xi}_\alpha$ for this section. To reproduce the equations of motion, it is enough to consider the Lagrangian at three points and vary with respect to the middle point \cite{lew_overview_2003}. Essentially, this works because discrete equations of motion propagate solutions time step by time step. One can discretize the Lagrangian in the following way,
\begin{equation}
\begin{aligned}
    \mathcal{L}\rightarrow &\mathcal{L}_d(\boldsymbol{\xi}^{n-1},\boldsymbol{\xi}^{n}, \phi^{n-1},\phi^{n}, \Delta t) \\ & +\mathcal{L}_d(\boldsymbol{\xi}^{n},\boldsymbol{\xi}^{n+1}, \phi^n,\phi^{n+1}, \Delta t).
\end{aligned}
\label{eq:threepointL}
\end{equation}

The equations of motion are then given through the discrete Euler-Lagrange equations. For example in the coordinate $\boldsymbol{\xi}$, this looks as follows,
\begin{equation}
    \begin{aligned}
        & \frac{d}{d\boldsymbol{\xi}^n} \mathcal{L}_d(\boldsymbol{\xi}^{n-1},\boldsymbol{\xi}^{n}, \phi^{n-1},\phi^{n}) \\ & - \frac{d}{d\boldsymbol{\xi}^n} \mathcal{L}_d(\boldsymbol{\xi}^{n},\boldsymbol{\xi}^{n+1}, \phi^n,\phi^{n+1}) = 0.
    \end{aligned}
	\label{eq:disc_euler_lagrange_eqn}
\end{equation}

For the most commonly used symplectic integrator - the leap-frog method - one chooses the piecewise Lagrangian in the following way \cite{lew_overview_2003},
\begin{equation}
\begin{aligned}
\mathcal{L}_d(\boldsymbol{\xi}^n,&\boldsymbol{\xi}^{n+1},\phi^n,\phi^{n+1}) \\ &= \mathcal{L}\left(\boldsymbol{\xi}^n, \frac{\boldsymbol{\xi}^{n+1}-\boldsymbol{\xi}^n}{\Delta t}, \phi(\boldsymbol{\xi}) \right),
\end{aligned}
\label{eq:leapfrog}
\end{equation}

where $\mathcal{L}$ is the time-continuous Lagrangian. Insertion into eq (\ref{eq:threepointL}) leads to the fully discretized three-point Lagrangian, from which the equations of motion can be obtained using the discrete Euler-Lagrange equations defined in eq  (\ref{eq:disc_euler_lagrange_eqn}).

\subsection{Final Lagrangian}
The step that remains is to combine the Ansatz for the P3M Lagrangian (\ref{eq:Lagrangian_Ansatz}) with the macroparticle decomposition (\ref{eq:macroparticles}), the spatial stepping formulae (\ref{eq:grid_potential}), (\ref{eq:discrete_field_term}) and the variational integrators (\ref{eq:threepointL}), (\ref{eq:disc_euler_lagrange_eqn}) in the Leap-Frog scheme (\ref{eq:leapfrog}).\ The fully temporally and spatially discrete three-point Lagrangian for the P3M method then reads as follows,
\begin{equation}
\begin{aligned}
\mathcal{L} = & \sum_\alpha \frac{m}{2}\omega_\alpha\left[\frac{\boldsymbol{\xi}_\alpha^{n}-\boldsymbol{\xi}_\alpha^{n-1}}{\Delta t}\right]^2 + \sum_\alpha \frac{m}{2}\omega_\alpha\left[\frac{\boldsymbol{\xi}_\alpha^{n+1}-\boldsymbol{\xi}_\alpha^n}{\Delta t}\right]^2 \\
& -\sum_\alpha q\omega_\alpha\int d^3\mathbf{x}\ (\phi^S)^{n-1}(\mathbf{x}) W^S(\mathbf{x}-\boldsymbol{\xi}^{n-1}_\alpha)
-\sum_\alpha q\omega_\alpha\sum_i(\phi^L)_i^{n-1} W^L(\mathbf{x}_i-\boldsymbol{\xi}^{n-1}_\alpha)  \\
&-\sum_\alpha q\omega_\alpha\int d^3\mathbf{x}\ (\phi^S)^{n}(\mathbf{x}) W^S(\mathbf{x}-\boldsymbol{\xi}^{n}_\alpha) -\sum_\alpha q\omega_\alpha\sum_i(\phi^L)_i^n W^L(\mathbf{x}_i-\mathbf{\xi}^n_\alpha) \\
& -\varepsilon_0h_xh_yh_z\sum_{i} \sum_{j} (\phi^L)_{i}^{n-1} (D_{xyz})_{i;j}(\phi^L)_{j}^{n-1} -\varepsilon_0\int d^3\mathbf{x}\ (\phi^S)^{n-1}(\mathbf{x}) \Delta (\phi^S)^{n-1}(\mathbf{x}) \\
& -\varepsilon_0h_xh_yh_z\sum_{i} \sum_{j} (\phi^L)_{i}^n (D_{xyz})_{i;j}(\phi^L)_{j}^n -\varepsilon_0\int d^3\mathbf{x}\ (\phi^S)^{n}(\mathbf{x}) \Delta (\phi^S)^{n}(\mathbf{x}).
\end{aligned}
\label{eq:full_p3m_lagrangian}
\end{equation}
The Lagrangian depends on sets of particle coordinates, continuous fields and sets of field values at grid points. If we use the discrete Euler-Lagrange equations (\ref{eq:disc_euler_lagrange_eqn}) with the above Lagrangian (\ref{eq:full_p3m_lagrangian}) for the variables $\boldsymbol{\xi}^k_\beta$, $(\phi^S)^k$, $(\phi^L)^k_m$ we obtain the equation of motion describing the $\beta$-th macroparticle, as well as a Poisson equation for each potential at the discrete time $k$,
\begin{subequations}
\begin{align}
\omega_\beta m \left[\frac{\boldsymbol{\xi}_{\beta}^{k+1}+\boldsymbol{\xi}_{\beta}^{k-1} - 2\boldsymbol{\xi}_{\beta}^{k}}{\Delta t^2}\right] &= -\sum_i \omega_\beta q_\beta \frac{dW^L}{d\xi}(\mathbf{x}_i-\boldsymbol{\xi}_\beta^k)(\phi^L)^k_i \label{eq:eom_p3m_coord} \\ & \quad - \omega_\beta q_\beta \int d^3\mathbf{x}\ \frac{dW^S}{d\xi}(\mathbf{x}-\boldsymbol{\xi}_\beta^k)(\phi^S)^k(\mathbf{x}) \notag  ,\\
\varepsilon_0h_xh_yh_z \sum_j (D_{xyz})_{m;j}(\phi^L)_j^k &= -\sum_\alpha \omega_\alpha q W^L(\mathbf{x}_m-\boldsymbol{\xi}_\alpha^k) \label{eq:eom_p3m_poiss_L} ,\\
\varepsilon_0\nabla^2 (\phi^S)^k(\mathbf{x}) &= -q\sum_\alpha \omega_\alpha q_\alpha W^S(\mathbf{x}-\boldsymbol{\xi}_\alpha^k) = [\rho^k - \rho^k\otimes\gamma](\mathbf{x}) . \label{eq:eom_p3m_poiss_S}
\end{align}
\end{subequations}

The volume factor $h_xh_yh_z$ in eq (\ref{eq:eom_p3m_poiss_L}) is due to assuming that the charge value at grid site $\mathbf{x}_i$ is the average over the cell. The derivative term of the interpolation function in eq (\ref{eq:eom_p3m_coord}) can be understood as a forward finite difference operator of first order if we use splines as above (\ref{eq:interpolation_splines_b0}),
\begin{align*}
\frac{d}{d\boldsymbol{\xi}} W(\mathbf{x}_i-\boldsymbol{\xi}) &= \frac{d}{d\boldsymbol{\xi}}\ b_1(\mathbf{x}_i-\boldsymbol{\xi}) \\
&= 3\frac{b_0(\mathbf{x}_{i+1}-\boldsymbol{\xi}) - b_0(\mathbf{x}_i-\boldsymbol{\xi})}{3h},
\end{align*}
where the factor 3 arises due to summation over three directions in the multi-index $i$. If summed over grid points, as in eq (\ref{eq:eom_p3m_coord}), this expression locates the grid points nearest to $\boldsymbol{\xi}$ and takes the finite difference of first order.

\section{Boundary Conditions}

The boundary condition enters the computation when Poisson's equations (\ref{eq:eom_p3m_poiss_L}), (\ref{eq:eom_p3m_poiss_S}) are solved. Since both potentials $\phi^S$, $\phi^L$ are solved differently we must distinguish between two Green's functions $G^L$, $G^S$ for both ranges and make use of the finite convolution $\otimes_d$ for the case of the long range interaction.

From eqns (\ref{eq:eom_p3m_poiss_L}), (\ref{eq:eom_p3m_poiss_S}), we see that $\rho^L$, $\rho^S$ can be expressed as,

\begin{subequations}
\begin{align}
	(\rho^L)^n_i &= -\sum_\alpha \omega_\alpha q W^L(\mathbf{x}_i-\boldsymbol{\xi}_\alpha^n)  \label{eq:rho_L} ,\\
	(\rho^S)^n &= -q\sum_\alpha \omega_\alpha q_\alpha W^S(\mathbf{x}-\boldsymbol{\xi}_\alpha^n), \label{eq:rho_S}
\end{align}
\end{subequations}

where again $n$ denotes a temporal index and $i$ a spatial multi-index. The solutions to eqns (\ref{eq:eom_p3m_poiss_L}), (\ref{eq:eom_p3m_poiss_S}) are then given as,
\begin{subequations}
\begin{align}
(\phi^L)_i^n &= \left[(\rho^L)^n\otimes_d G^L\right]_i \notag \\ &= \frac{h_x h_y h_z}{\varepsilon_0} \sum_q (\rho^L)^n_q G^L_{i;j} \label{eq:phi_L}\ \text{ and}\\ 
(\phi^S)^n(\mathbf{x}) &= ((\rho^S)^n\otimes G^S)(\mathbf{x}) \notag ,\\ 
&= \frac{1}{\varepsilon_0}\int d^3\mathbf{z}\ (\rho^S)^n(\mathbf{z})G^S(\mathbf{x}-\mathbf{z}) . \label{eq:phi_S}
\end{align}
\end{subequations}
Using spatial multi-indexes $i$ and $j$, the defining equations for Green's functions reads,
\begin{subequations}
\begin{align}
(D_{xyz})_{k}(G^L)_{i;j} &= \delta_{ij} , \label{eq:def_G_L}\\
\Delta_{\mathbf{x}} (G^S)(\mathbf{x},\mathbf{y}) &= \delta^{(3)}(\mathbf{x}-\mathbf{y}), \label{eq:def_G_S}
\end{align}
\end{subequations}

where $D_{xyz}$ represents the discrete Laplacian, as defined in eq (\ref{eq:3dim_second_order_fin_diff}). The discrete Laplacian appears since the long range problem is solved on a grid.

\subsection{Green's Function for a Planar Cathode}

For the purpose of quantifying the potential error we will consider the case of a planar grounded cathode. 
The unbounded problem, corresponding to $\phi(r)\overset{r\rightarrow\infty}{\longrightarrow} 0$, is solved by,
$$G^F(\mathbf{x}, \mathbf{y}) = \frac{1}{|\mathbf{x} - \mathbf{y}|}.$$

\begin{figure}[htp]
	\centering

	\begin{tikzpicture}[cross/.style={path picture={\draw[black] (path picture bounding box.south east) -- (path picture bounding box.north west) (path picture bounding box.south west) -- (path picture bounding box.north east);}}]
	\draw [black,->] (0,2.25) -- (1,2.25) ;
	\node [align=left] at (1.35,2.25) {$z$};
	\draw [black,->] (0,2.25) -- (0,3.25);
	\node [align=left] at (0.3,3.25) {$y$};
	\node [draw,circle,cross,minimum width=0.25 cm] at (0,2.25){};
    \node [align=left] at (-0.35,2.25) {$x$};

	\draw [ultra thick, black] (0,-2) -- (0,2);
	\draw [black] (-0.2,-1.8) to (-0.2,-1.8) node[ground]{};
	\draw [black] (0,-1.8) -- (-0.2,-1.8);
	
	\begin{scope}[xshift=-0.35cm, yshift=0.6cm]
	\node [shape=circle, minimum size=0.1cm, inner sep=0, fill=blue] at (2,0) {};
	\node [align=right] at (1.7, -0.1 ) {\small $\mathbf{y}$};
	\end{scope}
	
	\begin{scope}[xshift=0.35cm, yshift=0.6cm]
	\node [shape=circle, minimum size=0.1cm, inner sep=0, fill=black] at (- 2,0) {};
	\node [align=left] at (-2.5, -0.1) {\small $S(\mathbf{y})$};
	\end{scope}
	
	\draw [->] (1.65,0.6) -- (1.4, -0.4);
	\draw [->] (-1.65,0.6) -- (1.35, -0.425);
	\node [shape=circle, minimum size=0.1cm, inner sep=0, fill=black] at (1.4, -0.45) {};
	\node [align=left] at (1.6, -0.55) {$\mathbf{x}$};
	
	\node [cross, minimum width=0.25 cm] at (2,0) {};
	\node [align=left] at (2.35,0) {$\mathbf{r}_0$};
	
	\draw [color=red] (-3,-1) rectangle (-1,1);
	\draw[color=gray, fill=gray, opacity=0.3](-1,-1) rectangle (1,1);
	\draw[color=red] (1,-1) rectangle (3,1);
	

	\node [align=center] at (-2, -1.8) {\small \textcolor{red}{Mirror Charge} \\ \small \textcolor{red}{Domain}};
	\node [align=center] at (2, -1.8) {\small \textcolor{red}{Charge and} \\ \textcolor{red}{Field Domain}};
	\end{tikzpicture}
	\caption{Illustration of the planar Green's function. Decoupling the mirror charge from the field domain means, the (gray) space in between does not need to be resolved. $\mathbf{r}_o$ denotes the position of the centre of the charge and field domain, i.e. where the physical charges are and where the field is evaluated. 	\label{fig:planarGF}}
\end{figure}
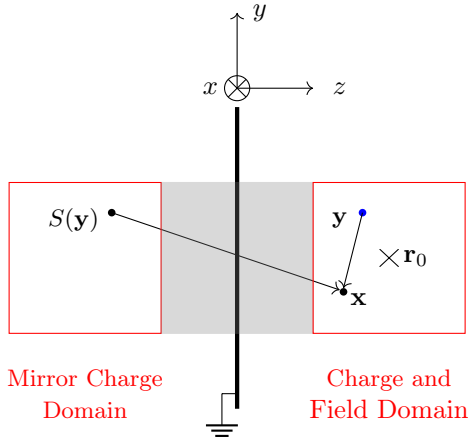

Let us consider the case of a planar cathode at $z=0$, which is depicted in figure \ref{fig:planarGF}. Green's function can be obtained by using mirror charges \cite{alshal_image_2018,grebenkov_geometrical_2013,nelson_using_1997}, which yields the following,

\begin{equation}
	G(\mathbf{x},\mathbf{y}) = G^F(\mathbf{x},\mathbf{y}) - G^F(\mathbf{x}, S(\mathbf{y})),
	\label{eq:GreenFunc_planar_cathode}
\end{equation}

where $S:\mathbb{R}^3\rightarrow\mathbb{R}^3$ is the mirror symmetry map,

\begin{equation}
	S(x,y,z) = (x,y,-z). 
	\label{eq:symmetry_planar_cathode}
\end{equation}

If we denote Green's function with $G(\mathbf{x}-\mathbf{y})$, as in (\ref{eq:GreenFunc_planar_cathode}), then $\mathbf{x}$ is the location where the field is calculated (field domain) and $\mathbf{y}$ is the location of the charges (charge domain). Numerically, the issue with eq (\ref{eq:GreenFunc_planar_cathode}) is that the location of the mirror image $S(\mathbf{y})$ (mirror charge domain) must be part of the computational domain as well. In other words, the entire distance between charge and image must be resolved as well (see figure \ref{fig:planarGF}). This is leads to poorer resolution at higher computational cost.

As demonstrated by \textcite{qiang_parallel_2004}, one can reduce the work load by shifting Green's function by the centre of the charge domain. This method decouples the location of the image (mirror charge domain) from the location where its field is calculated (charge and field domain). If Green's function is denoted as $G(\mathbf{x}, \mathbf{y})$, then $\mathbf{x}$ lies in the field domain and $\mathbf{y}$ in the charge domain. If we denote with $\mathbf{r}_o$ the location of the original charge, we can shift Green's function, which reduces the domain needed for $\mathbf{x}$. The resulting  Green's function is then given as,
\[
G(\mathbf{x},\mathbf{y}) = G^F(\mathbf{x},\mathbf{r}_o-\mathbf{y}) - G^F(\mathbf{x},\mathbf{r}_o- S(\mathbf{y})).
\]
\subsubsection{Short Range Potential Error}

Let $\Phi$ be the true potential that is to be approximated. The charges, from which the potential originates, may or may not be in the proximity of conductors, such as a photocathode or beam tubes, which impose electrostatic boundary conditions. Without specifying the geometry of the conductors, let us denote the corresponding Green's function as $R$, assuming we know its true form. For example in the absence of conductors $R:=G^F$ would be the true form. Given this notation, $\Phi$ can be written as the convolution between the true Green's function $R$ for the chosen geometry and the total continuous charge distribution,
\begin{equation*}
\begin{aligned}
    \Phi(\mathbf{x}) &= R\otimes\rho \\ &= \frac{q}{\varepsilon_0}\int d^3\mathbf{y}\ R(\mathbf{x},\mathbf{y})\sum_\alpha\omega_\alpha S_x(\mathbf{y}-\boldsymbol{\xi}_\alpha^n),
\end{aligned}
\end{equation*}
where eq (\ref{eq:macropart_charge_dists}) was used.

The total computed potential using P3M will be denoted as $\phi^\text{tot}$. It can be obtained by adding eqns (\ref{eq:phi_L}) and (\ref{eq:phi_S}) and insert from eqns (\ref{eq:rho_L}), (\ref{eq:rho_S}). Additionally, the long-range part needs to be re-interpolated, using $W^L$. For a more general picture, we can remove the fixed charge distribution and instead write the dependence on the charge position explicitly. The potentials depend on charge position $\boldsymbol{\xi}_\alpha$ and are evaluated at $\mathbf{x}$, i.e. they determine the potential value at $\mathbf{x}$ due to one particle at $\boldsymbol{\xi}_\alpha$. For the sake of simplicity, we set $\omega_\alpha=1$, so the potentials are given as,

\begin{subequations}
\begin{align}
    \Phi(\mathbf{x},\boldsymbol{\xi}_\alpha^n) &= \frac{q}{\varepsilon_0}\int d^3\mathbf{y}\ R(\mathbf{x},\mathbf{y}) S_x(\mathbf{y}-\boldsymbol{\xi}_\alpha^n) \label{eq:true_phi_def} \\
\phi^\text{tot}(\mathbf{x}, \boldsymbol{\xi}) &= \frac{q}{\varepsilon_0}\int d^3\mathbf{y}\ G^S(\mathbf{x},\mathbf{y}) W^S(\mathbf{y}-\boldsymbol{\xi}) \notag \\ &\quad + \frac{q}{\varepsilon_0}\sum_{i,j} \cdot W^L(\mathbf{x}_i-\boldsymbol{\xi})G^L_{i;j}
\notag \\ &\qquad\qquad\quad W^L\left(\mathbf{x}-\mathbf{x}_i\right) \label{eq:phi_tot_def}
\end{align}
\end{subequations}

A statement about the overall error of this method is difficult to make, as there are interdependent contributions from discretization, interpolation and finite differences. In addition the difference between $\phi^\text{tot}$ and $\Phi$ depends on the position and the amount of charges present. An error measure that captures the overall deviation can be found by integrating, as used by \textcite{hockney_computer_1988}, over both degrees of freedom. Let us denote the error $E$ by,
\begin{equation}
E = \int d^3\boldsymbol{\xi}_\alpha^n\int d^3\mathbf{x}\ |\phi^\text{tot}(\mathbf{x},\boldsymbol{\xi}_\alpha^n) -\Phi(\mathbf{x},\boldsymbol{\xi}_\alpha^n)|^2
\label{eq:error_measure}
\end{equation}

In order to quantify $E$, made by summing the short-range contributions with the free interaction potential, we can assume that the total error is dominated by the short-range error. If the long-range part approximates the continuous case well, then this assumption holds. It will be proven afterwards that there is an expression for $G^L$, such that the deviation is indeed minimized. 
With this in mind, the error reads,
\begin{equation}
E = \int d^3\boldsymbol{\xi}_\alpha^n\int d^3\mathbf{x}\ |\phi^S(\mathbf{x},\boldsymbol{\xi}_\alpha^n) -\Phi^S(\mathbf{x},\boldsymbol{\xi}_\alpha^n)|^2,
\label{eq:short_range_error}
\end{equation}
where,
\begin{equation}
    \begin{aligned}
    &\Phi^S(\mathbf{x},\boldsymbol{\xi}_\alpha^n) \\ &= q\int d^3\mathbf{y}\ R(\mathbf{x},\mathbf{y}) (S_x-S_x\otimes\gamma)(\mathbf{y}-\boldsymbol{\xi}_\alpha^n) \\ &= q\int d^3\mathbf{y}\ G^S(\mathbf{x},\mathbf{y}) W^S(\mathbf{y}-\boldsymbol{\xi}_\alpha^n),
    \end{aligned}
    \label{eq:phi_S_def}
\end{equation}
denotes the true short-range potential. We again assume the case of a planar cathode, which means that $R$ is given by eq (\ref{eq:GreenFunc_planar_cathode}). Furthermore, we can consider point charges represented by a delta function and get
\begin{align*}
E &= \int d^3\mathbf{x}\int d^3\boldsymbol{\xi}_\alpha^n\ \left|\int d^3\mathbf{y}\ \left[ W^S(\mathbf{y}-\boldsymbol{\xi}_\alpha^n)\right]\left(\frac{1}{|\mathbf{x}-\mathbf{y}|} - R(\mathbf{x},\mathbf{y})\right)\right|^2 \\
&= \int d^3\mathbf{x}\int d^3\boldsymbol{\xi}_\alpha^n\ \left|\int d^3\mathbf{y}\ \left[ W^S(\mathbf{y}-\boldsymbol{\xi}_\alpha^n)\right]\frac{1}{|\mathbf{x}-S(\mathbf{y})|}\right|^2 \\
&= \int d^3\mathbf{x}\int d^3\boldsymbol{\xi}_\alpha^n\ \left|\int d^3\mathbf{y}\ \left[S_x(\mathbf{y}-\boldsymbol{\xi}_\alpha^n)-(S_x\otimes\gamma)(\mathbf{y}-\boldsymbol{\xi}_\alpha^n)\right]\frac{1}{|\mathbf{x}-S(\mathbf{y})|}\right|^2 \\
&= \int d^3\mathbf{x}\int d^3\boldsymbol{\xi}_\alpha^n\ \left|\int d^3\mathbf{y}\ \left[\delta(\mathbf{y}-\boldsymbol{\xi}_\alpha^n) - \gamma(\mathbf{y}-\boldsymbol{\xi}_\alpha^n)\right]\frac{1}{|\mathbf{x}-S(\mathbf{y})|}\right|^2 \\
&= \int d^3\mathbf{x}\int d^3\boldsymbol{\xi}_\alpha^n\ \left|\frac{1}{|\mathbf{x}-S(\boldsymbol{\xi}_\alpha^n)|} - \int d^3\mathbf{y} \frac{\gamma(\mathbf{y}-\boldsymbol{\xi}_\alpha^n)}{|\mathbf{x}-S(\mathbf{y})|} \right|^2.
\end{align*}

The action of the mirror charge onto the original charge location is equivalent to the action of the charge to the mirror charge domain. Hence, we can rewrite the expression for the error as
\begin{align*}
E &= \int d^3\mathbf{x}\int d^3\boldsymbol{\xi}_\alpha^n\ \left|\frac{1}{|S(\mathbf{x})-\boldsymbol{\xi}_\alpha^n|} - \int d^3\mathbf{y} \frac{\gamma(\mathbf{y}-\boldsymbol{\xi}_\alpha^n)}{|S(\mathbf{x})-\mathbf{y}|} \right|^2 \\
&= \int d^3\mathbf{x}\int d^3\boldsymbol{\xi}_\alpha^n\ \left|\frac{1}{|S(\mathbf{x})-\boldsymbol{\xi}_\alpha^n|} - \int d^3\mathbf{y} \frac{\gamma(\mathbf{y})}{|(S(\mathbf{x})-\boldsymbol{\xi}_\alpha^n)-\mathbf{y}|} \right|^2 \\
&= \int d^3\mathbf{x}\int d^3\boldsymbol{\xi}_\alpha^n\ \left|\frac{\text{Erfc}(\alpha|S(\mathbf{x})-\boldsymbol{\xi}_\alpha^n|)}{|S(\mathbf{x})-\boldsymbol{\xi}_\alpha^n|} \right|^2 \\
&= \int d^3\mathbf{x}\int d^3\boldsymbol{\xi}_\alpha^n\ \left|\frac{\text{Erfc}(\alpha|\mathbf{x}-S(\boldsymbol{\xi}_\alpha^n)|)}{|\mathbf{x}-S(\boldsymbol{\xi}_\alpha^n)|} \right|^2 \\
&\leq \left|\int d^3\mathbf{x}\int d^3\boldsymbol{\xi}_\alpha^n\ \frac{\text{Erfc}(\alpha|\mathbf{x}-S(\boldsymbol{\xi}_\alpha^n)|)}{|\mathbf{x}-S(\boldsymbol{\xi}_\alpha^n)|} \right|^2 .\\
\end{align*}

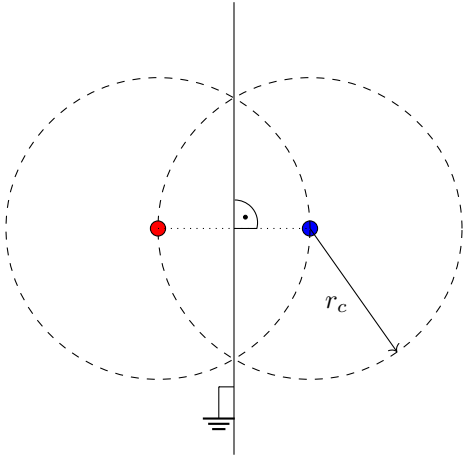
\begin{figure}[htp]
	\centering
	\begin{tikzpicture}
	\draw (0,-3) -- (0,3);
	\draw [fill = blue] (1,0) circle [radius=0.1cm];
	\draw [dashed] (1,0) circle [radius=2cm];
	\draw [fill = red] (-1,0) circle [radius=0.1cm];
	\draw [dashed] (-1,0) circle [radius=2cm];
	\draw [->, rotate around={35:(1,0)}] (1,0) -- (1,-2);
	\node[align=left] at (1.35,-1) {$r_c$};
	\draw [dotted] (-1,0) -- (1,0);
	\draw [black] (-0.2,-2.1) to (-0.2,-2.1) node[ground]{};
	\draw [black] (0,-2.1) -- (-0.2,-2.1);
	\draw (0,0) -- (0:0.3cm) arc (-15:90:0.3cm);
	\draw [fill=black] (0.15,0.15) circle [radius=0.025];
	\end{tikzpicture}
	\caption{When $r_c$ denotes the cut-off radius of the method, only particles closer than $\frac{r_c}{2}$ to the cathode are affected by the boundary condition.}
	\label{fig:ShortError}
\end{figure}

As a consequence, the error is equal to the integral of the short-range potential from the corresponding mirror charge. Due to the strong decay of the complementary error function, the error is strongly confined. If we restrict the integration to the original charge domain and require that the image charge is sufficiently far away from the original, the error is nil. Consequentially, only if a particle is within half the Ewald cut-off length to the cathode, the error (within the computational domain) will be non-vanishing as can be seen in figure \ref{fig:ShortError}.

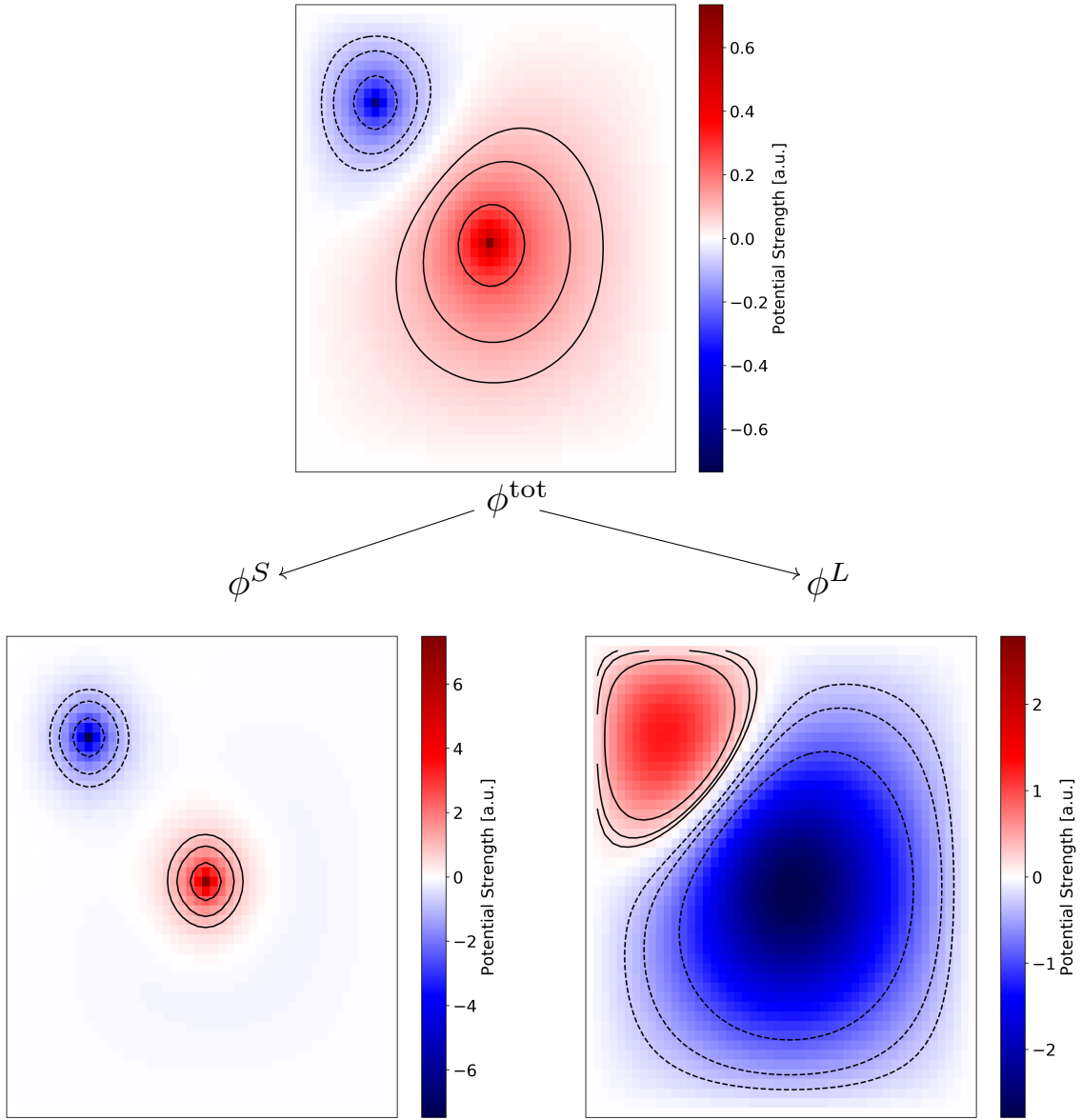
\begin{figure}[htbp]
\centering
    \begin{tikzpicture}
    \node [anchor=south west,inner sep=0] at (0,0) {\includegraphics[width=7cm]{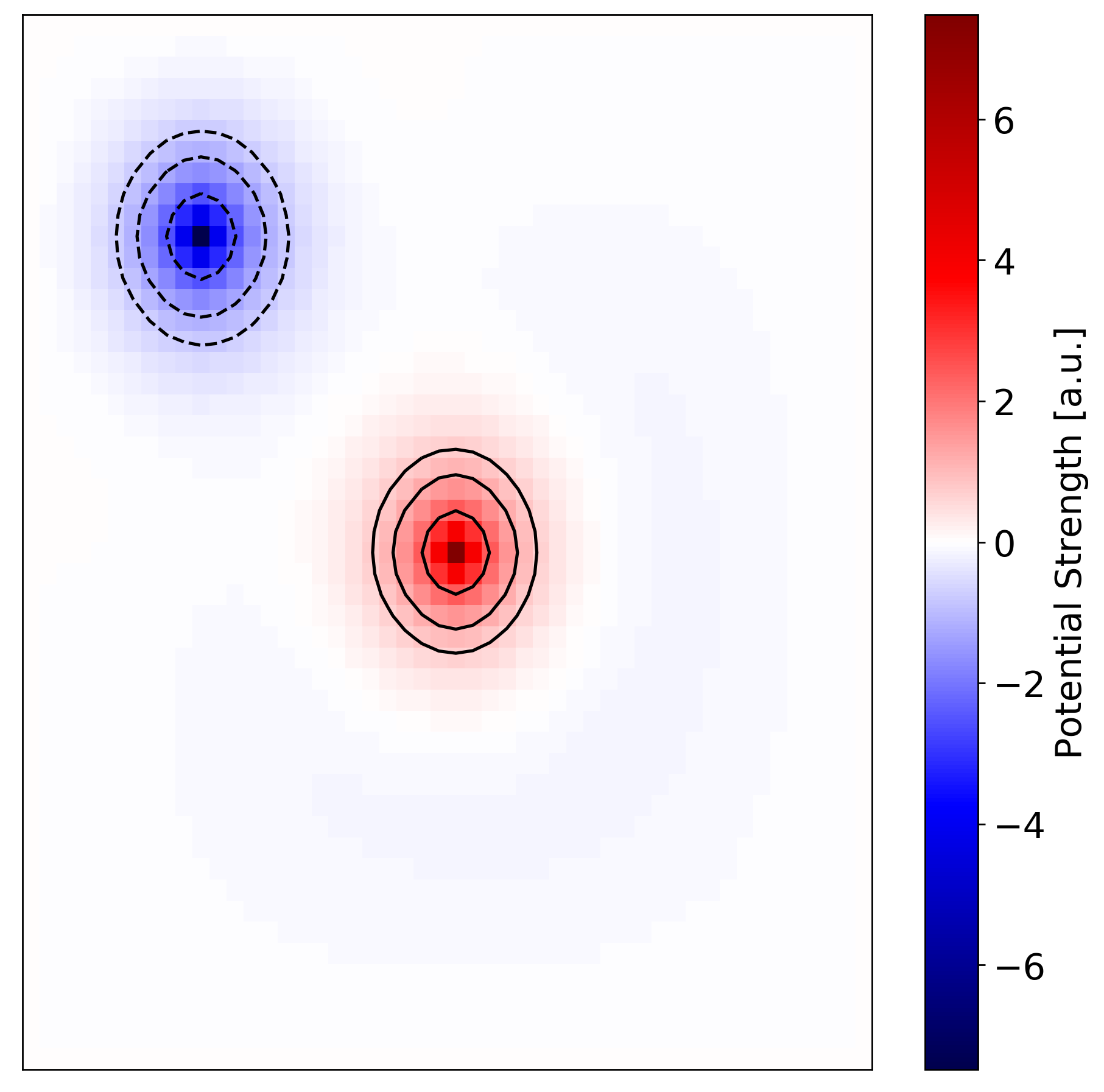}};
    
    \node [anchor=south west,inner sep=0] at (8,0) {\includegraphics[width=7cm]{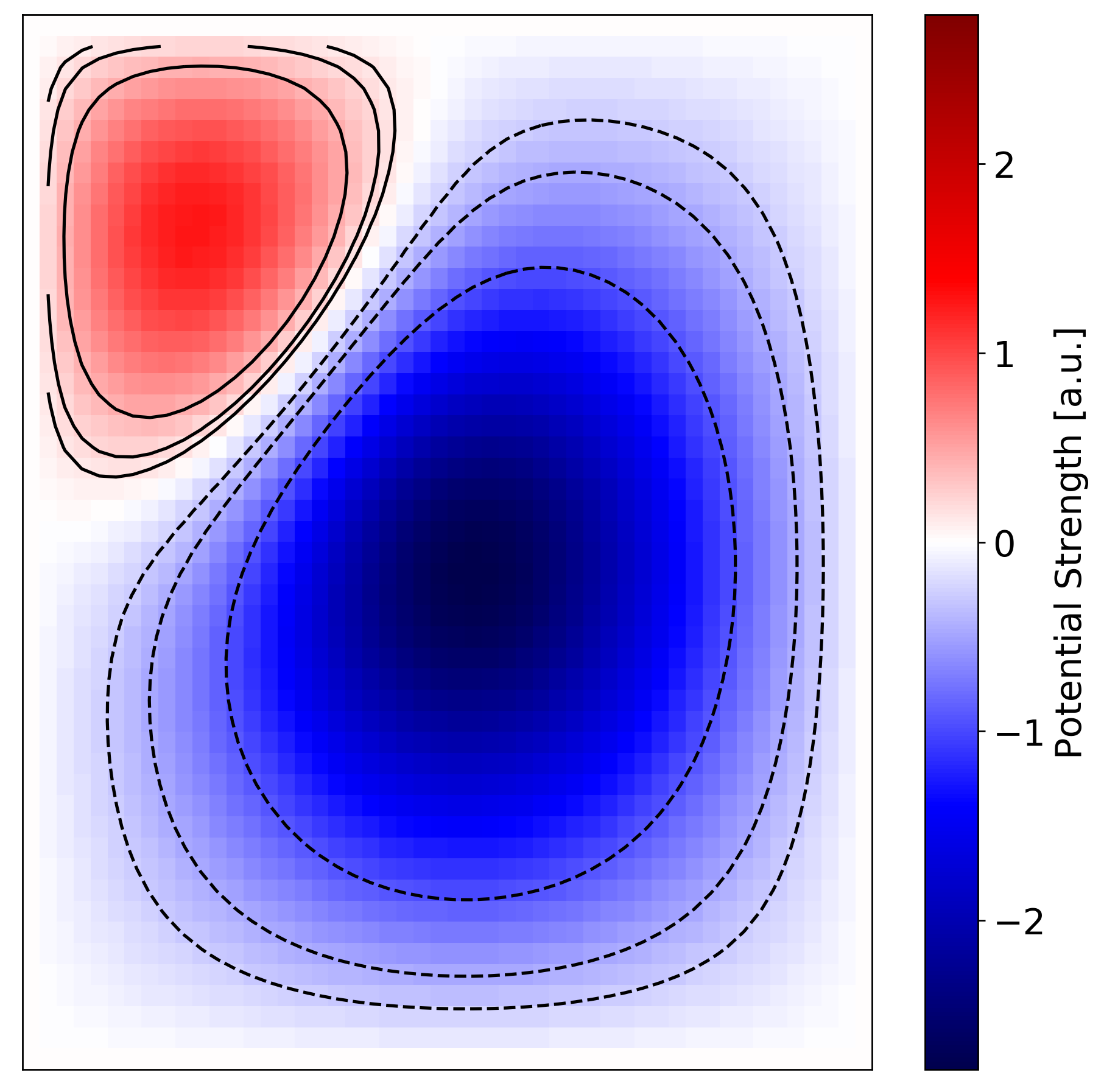}};
    
    \node [anchor=south west,inner sep=0] at (4,9) {\includegraphics[width=7cm]{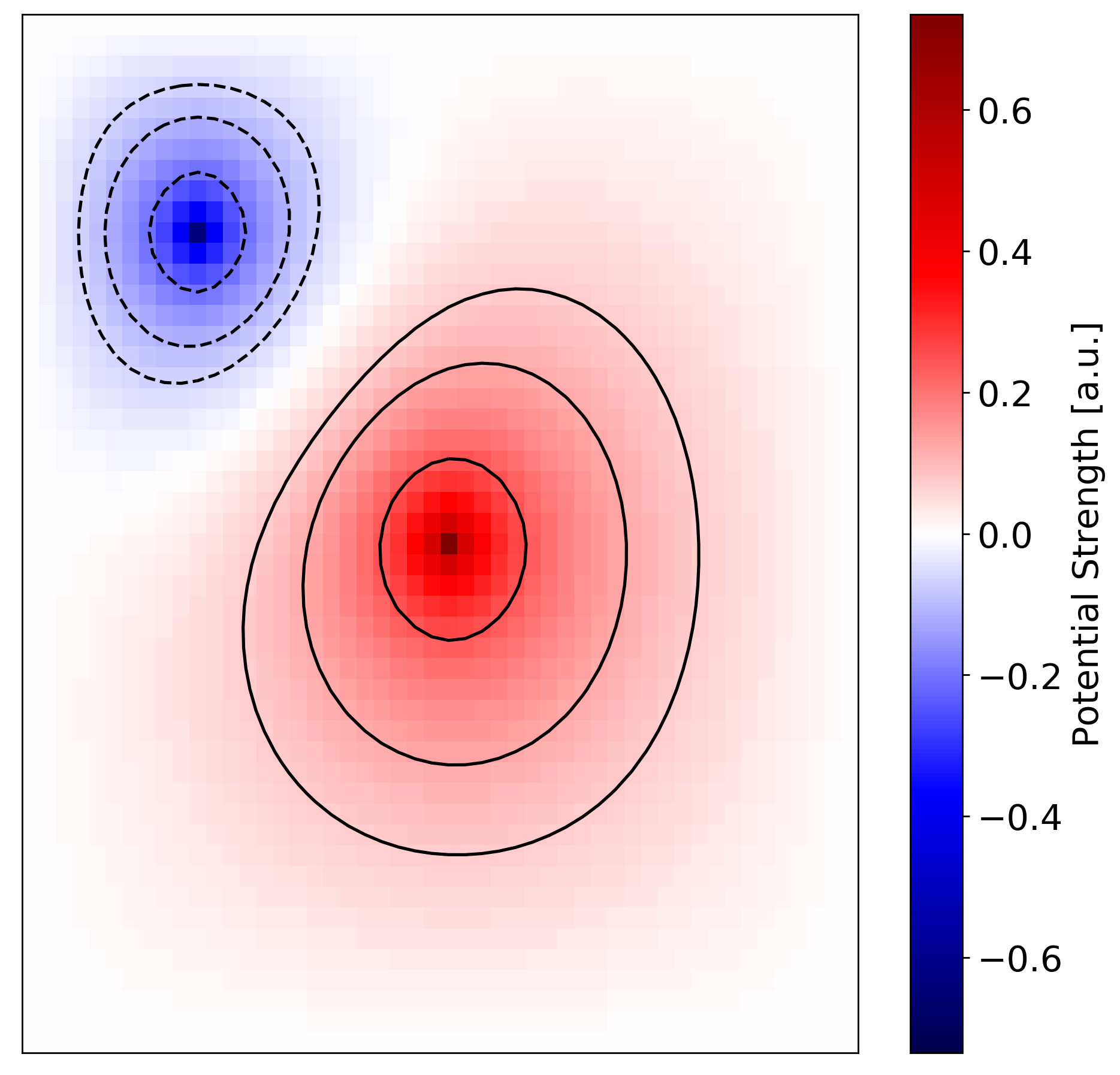}};
    
    \node [align=left] at (7.2,8.75) {\scalebox{1.5}{$\phi^\text{tot}$}};
    \node [align=left] at (3.5,7.6) {\scalebox{1.5}{$\phi^S$}};
    \node [align=left] at (11.5, 7.6) {\scalebox{1.5}{$\phi^L$}};
    
    \draw [->] (6.6,8.6) -- (3.9,7.7);
    
    \draw [->] (7.5,8.6) -- (11.1,7.7);
    
    \end{tikzpicture}
    \caption{Visualization of the three potentials - in  coulomb/meter - for two oppositely charged particles in a grounded box. The potentials were computed by solving Poisson's equations $ \Delta\phi^\text{tot} = \rho$, $ \Delta \phi^S = \rho^S$ and $\Delta \phi^L = \rho^L$  using a Gauss Seidel solver on a grid of length 50. The charge distributions $\rho$, $\rho^S$ and $\rho^L$ are understood as defined in eq (\ref{eq:charge_splitting}). The dotted and solid lines represent contour lines of the potential.}
    \label{fig:boundary_gauss_seidel_results}
\end{figure}

For any other geometry, we may infer that although the Green's function will be different, the result will be the same, as it arises due to the charge screening, which is independent of the geometry. By solving both Poisson's equations with the same boundary conditions, we essentially solve the same problem twice, but on different length scales. Instead, one can solve both equations separately, with and without a boundary condition respectively. The error, as has been shown above, is negligible. A visualization of the short-ranged potential can be seen in figure \ref{fig:boundary_gauss_seidel_results}, which illustrates that boundary conditions have no impact in the short range. For the long-ranged potential the boundaries do influence the potential.


\subsubsection{Long Range Potential Error}

The following derivation closely follows the derivation of Hockney \& Eastwood \cite{hockney_computer_1988} for the optimal influence function.

In the above calculation, it was assumed that the long-range \textit{computed} potential approximates reality, and that the error was predominantly coming from the short-ranged part, as written in eq (\ref{eq:short_range_error}). The existence of such an approximation will be shown in the following section.
With regard to optimizing the long-range part, the calculations become easier if we use $G^S=R$. The optimization condition reads as follows,
\begin{equation}
\label{eq:minimization_condition}
\begin{aligned}
\frac{\delta}{\delta G^L_{k;q}} & \int d^3\boldsymbol{\xi}_\alpha^n\int d^3\mathbf{x}\ |\phi^\text{tot}(\mathbf{x},\boldsymbol{\xi}_\alpha^n) -\Phi(\mathbf{x},\boldsymbol{\xi}_\alpha^n)|^2  \\ & \overset{!}{=} 0,
\end{aligned}
\end{equation}
where $\delta E/\delta G$ represents a functional derivative, as it appears e.g. in Hamilton's principle and the expressions $\Phi$ and $\phi^\text{tot}$ have been defined in eqns (\ref{eq:true_phi_def}) and (\ref{eq:phi_tot_def}). Again, $G^L_{k;q}$ represents Green's function in the long range, as defined in eq (\ref{eq:def_G_L}).

The general form of the potentials, the fact that $\phi^\text{tot}$ is partially discrete and the many integrations makes this derivative difficult to evaluate, luckily one can simplify the calculation tremendously by transforming to Fourier space. For the non-discrete parts of eqns (\ref{eq:true_phi_def}) and (\ref{eq:phi_tot_def}), this poses no challenge. If we denote a Fourier space vector with $\mathbf{k}$ the transforms read,

\begin{subequations}
\begin{align}
\hat{\Phi}(\mathbf{k}) &= q\frac{1}{\varepsilon_0}\sum_\alpha\omega_\alpha \hat{R}(\mathbf{k})\hat{S}_x(\mathbf{k})e^{-i\mathbf{k}\boldsymbol{\xi}_\alpha^n}, \label{eq:True_potential_FT} \\
\hat{\phi}^S(\mathbf{k}) &= \frac{1}{\varepsilon_0}\sum_\alpha\omega_\alpha\widehat{W}^S(\mathbf{k})\hat{R}(\mathbf{k})e^{-i\mathbf{k}\boldsymbol{\xi}_\alpha^n}. \label{eq:phi_S_FT}
\end{align}
\end{subequations}

The long-range potential eq (\ref{eq:phi_L}) is the result of a discrete convolution, which fourier-transforms to the product of Fourier series coefficients. In order to be able to compare to the continuous transforms, we rewrite the discrete convolution using a \textit{Dirac comb},
$$\Sha_{h_x}(x) = \sum_{m=-\infty}^\infty \delta(x-mh_x).$$
In three dimensions this generalizes to the corresponding product,
$$\Sha(\mathbf{x}) = \Sha_{h_x}(x)\Sha_{h_y}(y)\Sha_{h_z}(z).$$
The corresponding Fourier transform is a Dirac comb with the period of the reciprocal lattice,
$$\widehat{\Sha}_{2\pi/h_x}(k_x) = \sum_{m=-\infty}^\infty \delta(k_x-m2\pi/h_x).$$
Using this, the discrete convolution in eq (\ref{eq:phi_L}) can be written in continuous form, assuming we know the continuous form of $G^L$ and $W^L$, which is usually the case but can be readily obtained by making the grid constants infinitely small \cite{hollos_lattice_2005}.\ The result is a continuous function $(\tilde{\phi}^L)^n$ that reproduces the meshed potential at grid points $\mathbf{x}_k$,
\begin{align*}
(\tilde{\phi}^L)^n(\mathbf{x}) &= \int d^3\mathbf{y}\ W^L(\mathbf{y}-\boldsymbol{\xi}^n_\alpha)\Sha(\mathbf{y}) \\ & \qquad\qquad\quad \cdot G^L(\mathbf{x},\mathbf{y})\\
(\tilde{\phi}^L)^n(\mathbf{x}_k) &= (\phi^L)^n_k.
\end{align*}

The re-interpolated long-range potential is given as the convolution with the interpolation function \cite{hockney_computer_1988},
\begin{equation}
	(\phi^L)^n(\mathbf{x}) = (W\otimes (\tilde{\phi}^L)^n)(\mathbf{x}).
	\label{eq:cont_conv_L_pot}
\end{equation}

The continuous Fourier transform of (\ref{eq:cont_conv_L_pot}) can then be readily calculated,
$$\widehat{(\tilde{\phi}^L)^n}(\mathbf{k}) = (\widehat{W}^L\otimes\Sha_{(2\pi/h)})(\mathbf{k})\widehat{G}e^{-i\mathbf{k}\boldsymbol{\xi}^n_\alpha}.$$

The convolution can be evaluated directly. Let $\mathbf{m}\in\mathbb{N}^3$ denote a vector of integers,
\begin{align*}
& \widehat{W}^L\otimes\widehat{\Sha}_{(2\pi/h)}(\mathbf{k}) \\ &= \int d^3\ \mathbf{q}\ \sum_{\mathbf{m}} \delta(\mathbf{k}_x-{m}_x2\pi/h_x) \\ & \qquad\qquad\qquad\ \cdot \delta(\mathbf{k}_y-{m}_y2\pi/h_y)\\ & \qquad\qquad\qquad\ \cdot \delta(\mathbf{k}_z-{m}_z2\pi/h_z) W(\mathbf{k}) \\
&= \sum_{\mathbf{m}} \widehat{W}^L\left[\mathbf{k} - 2\pi(m_x/h_x, m_y/h_y, m_z/h_z)\right] \\
&=: \sum_{\mathbf{m}} \widehat{W}^L\left[\mathbf{k} - \frac{2\pi}{h}\mathbf{m}\right].
\end{align*}

The sum can be identified as aliasing sum \cite{hockney_computer_1988}. It describes the influence of the aliases of the principal harmonic of the charge distribution, i.e. every harmonic that is periodically shifted to the original. Basically, it arises due to the sampling onto the mesh that is done during the charge assignment. 

The corresponding continuous Fourier representation of the long-range potential reads as follows,
\begin{align*}
(\widehat{\phi^L})^n(\mathbf{k}) &= \widehat{W}^L(\mathbf{k})\widehat{(\tilde{\phi}^L)}^n(\mathbf{k}) \\
&= \widehat{W}^L(\mathbf{k})\hat{G}(\mathbf{k})\sum_{\mathbf{m}}\widehat{W}^L\left[\mathbf{k} - \frac{2\pi}{h}\mathbf{m}\right] e^{-i\mathbf{k}\boldsymbol{\xi}^n_\alpha}.
\end{align*}


The total potential thus has the following form,
\begin{align*}
\phi^\text{tot}(\mathbf{x}) &= \sum_\alpha \sum_{\mathbf{k}} \left[\widehat{W}^S(\mathbf{k})\widehat{G}^S(\mathbf{k})e^{i\mathbf{k}\boldsymbol{\xi}_\alpha^n} + \widehat{W}^L(\mathbf{k})\hat{G}(\mathbf{k})\sum_{\mathbf{m}}\widehat{W}^L\left[\mathbf{k} - \frac{2\pi}{h}\mathbf{m}\right]e^{-i(\mathbf{k}-2\pi\mathbf{n}/h)\boldsymbol{\xi}^n_\alpha} \right. \\ & \qquad\qquad\quad -\hat{R}(\mathbf{k})\hat{S}_x(\mathbf{k})e^{i\mathbf{k}\boldsymbol{\xi}_\alpha^n}\biggr] e^{-i\mathbf{k}\mathbf{x}}.
\end{align*}

By omitting the the sum over particles $\alpha$, we can again rewrite this to the potential caused by one particle at position $\boldsymbol{\xi}_\alpha$,

\begin{align*}
\phi^\text{tot}(\mathbf{x},\boldsymbol{\xi}_\alpha^n) &= \sum_{\mathbf{k}} \biggl[\widehat{W}^S(\mathbf{k})\widehat{G}^S(\mathbf{k})e^{i\mathbf{k}\boldsymbol{\xi}_\alpha^n} + \widehat{W}^L(\mathbf{k})\hat{G}(\mathbf{k})\sum_{\mathbf{m}}\widehat{W}^L\left[\mathbf{k} - \frac{2\pi}{h}\mathbf{m}\right]e^{-i(\mathbf{k}-2\pi\mathbf{n}/h)\boldsymbol{\xi}_\alpha^n}  \\ & \qquad\qquad -\hat{R}(\mathbf{k})\hat{S}_x(\mathbf{k})e^{i\mathbf{k}\boldsymbol{\xi}_\alpha^n}\biggr] e^{-i\mathbf{k}\mathbf{x}}.
\end{align*}

In addition, potentials depend only on absolute distance values, hence it makes sense to write a dependence on the particle position and the distance from it. We define $\mathbf{r}=\mathbf{x}-\boldsymbol{\xi}^n$, which yields,

\begin{align*}
& \phi^\text{tot}(\mathbf{r},\boldsymbol{\xi}_\alpha^n) = \sum_{\mathbf{k}} \biggl[\widehat{W}^S(\mathbf{k})\widehat{G}^S(\mathbf{k})e^{i\mathbf{k}\boldsymbol{\xi}_\alpha^n} + \widehat{W}^L(\mathbf{k})\hat{G}(\mathbf{k})\sum_{\mathbf{m}}\widehat{W}^L\left[\mathbf{k} - \frac{2\pi}{h}\mathbf{m}\right]e^{-i(\mathbf{k}-2\pi\mathbf{n}/h)\boldsymbol{\xi}_\alpha^n + i\mathbf{k}\boldsymbol{\xi}_\alpha^n} \\ & \qquad\qquad\qquad\qquad\quad -\hat{R}(\mathbf{k})\hat{S}_x(\mathbf{k})e^{i\mathbf{k}\boldsymbol{\xi}^n}\biggr] e^{-i\mathbf{k}\mathbf{r}}.
\end{align*}

The Fourier transform can be read off as

\begin{align*}
& \widehat{\phi}^\text{tot}(\mathbf{k},\boldsymbol{\xi}_\alpha^n) = \biggl[\widehat{W}^S(\mathbf{k})\widehat{G}^S(\mathbf{k})e^{i\mathbf{k}\boldsymbol{\xi}_\alpha^n} + \widehat{W}^L(\mathbf{k})\hat{G}(\mathbf{k})\sum_{\mathbf{m}}\widehat{W}^L\left[\mathbf{k} - \frac{2\pi}{h}\mathbf{m}\right]e^{-i(\mathbf{k}-2\pi\mathbf{n}/h)\boldsymbol{\xi}_\alpha^n + i\mathbf{k}\boldsymbol{\xi}_\alpha^n} \\ & \qquad\qquad\qquad\qquad\quad  -\hat{R}(\mathbf{k})\hat{S}_x(\mathbf{k})e^{i\mathbf{k}\boldsymbol{\xi}_\alpha^n}\biggr].
\end{align*}

We can then propose the following minimization condition analogous to the above, while remembering that $\Phi$ and $\phi^S$ are independent of $G^L$:

\begin{align} \label{eq:minE}
    \hat{E} &= \frac{\delta}{\delta (\widehat{G}^L)(\mathbf{k})} \int d^3\boldsymbol{\xi}_\alpha^n\ \left|\widehat{\phi}^\text{tot}(\mathbf{k},\boldsymbol{\xi}_\alpha^n)-\hat{\Phi}(\mathbf{k})\right|^2 = \int d^3\boldsymbol{\xi}_\alpha^n\ 2\left|\widehat{\phi}^\text{tot}(\mathbf{k},\boldsymbol{\xi}_\alpha^n)-\hat{\Phi}(\mathbf{k})\right| \frac{\delta \widehat{\phi}^L(\mathbf{k},\boldsymbol{\xi}_\alpha^n)}{\delta (\widehat{G}^L)(\mathbf{k})} \overset{!}{=} 0 .    
\end{align}

The integral over $\boldsymbol{\xi}_\alpha$ can be resolved to obtain $\delta(\mathbf{k}- (\mathbf{k}-2\pi\mathbf{n}/h))$ and solve for $\widehat{G}^L$, while keeping in mind that $\widehat{G}^L$ is a periodic function in $\mathbf{k}$-space. For simplicity, we will set $\widehat{G}^S = \hat{R}$ to get the following,

\begin{align*}
\hat{E} &= \sum_{\mathbf{m}}(\widehat{W}^L)^2\left[\mathbf{k} - \frac{2\pi}{h}\mathbf{m}\right] \left[ \widehat{G}^L(\mathbf{k})\sum_{\mathbf{m}}(\widehat{W}^L)^2\left[\mathbf{k} - \frac{2\pi}{h}\mathbf{m}\right] -\hat{R}\left[\mathbf{k} - \frac{2\pi}{h}\mathbf{m}\right](\hat{S}_x-\widehat{W}^S)\left[\mathbf{k} - \frac{2\pi}{h}\mathbf{m}\right]\right]  \\
&= \widehat{G}^L(\mathbf{k})\left|\sum_{\mathbf{m}}(\widehat{W}^L)^2\left[\mathbf{k} - \frac{2\pi}{h}\mathbf{m}\right]\right|^2 - \sum_{\mathbf{m}}(\widehat{W}^L)^2\left[\mathbf{k} - \frac{2\pi}{h}\mathbf{m}\right]\hat{R}\left[\mathbf{k} - \frac{2\pi}{h}\mathbf{m}\right](\hat{S}_x-\widehat{W}^S)\left[\mathbf{k} - \frac{2\pi}{h}\mathbf{m}\right] \\
&\overset{!}{=} 0.
\end{align*}

This equation can be solved for $\widehat{G}^L(\mathbf{k})$, which leads to the following expression,
\begin{equation}\label{eq:Ghfinal}
\widehat{G}^L(\mathbf{k}) = \frac{\sum_{\mathbf{m}}(\widehat{W}^L)^2\left[\mathbf{k} - \frac{2\pi}{h}\mathbf{m}\right]\hat{R}\left[\mathbf{k} - \frac{2\pi}{h}\mathbf{m}\right](\hat{S}_x-\widehat{W}^S)\left[\mathbf{k} - \frac{2\pi}{h}\mathbf{m}\right]}{\left|\sum_{\mathbf{m}}(\widehat{W}^L)^2\left[\mathbf{k} - \frac{2\pi}{h}\mathbf{m}\right]\right|^2}.
\end{equation}

This result eq (\ref{eq:minE}) solves the minimization condition eq (\ref{eq:minimization_condition}) and shows, that the minimum assumed previously exists. In the following we will show a way to simplify the unwieldy expression eq (\ref{eq:minE}) in order to be more suitable for application.
Per definition (\ref{eq:WL}), (\ref{eq:WS}), the Fourier transforms of the interpolation functions, are given as follows,
\begin{align*}
\widehat{W}^S &= \hat{S}_x - \hat{S}_x\hat{\gamma} \\
\widehat{W}^L &= \hat{S}_x\hat{\gamma}\hat{b}_0,
\end{align*}

therefore all aliasing sums contain a decaying factor $\hat{\gamma}(\mathbf{k}) = e^{-k^2/(4\alpha^2)}$. If mesh-size and Ewald-sphere are chosen sufficiently small \cite{deserno_how_1998}, we can only keep the principal harmonic $\mathbf{m}=0$. Equation (\ref{eq:Ghfinal}) can then be rewritten as
$$\widehat{G}^L(\mathbf{k}) = \frac{\hat{R}\hat{S}_x\hat{\gamma}}{\left(\hat{S}_x\hat{\gamma}\hat{b}_0\right)^2} = \hat{R}\cdot\frac{1}{\hat{\gamma}\hat{S}_x\hat{b}_0^2},$$
which provides a handy formula for the optimal Green's function.

\section{Conclusions}


Mesh-based algorithms are commonly employed in the simulation of charged many-body systems, since typically, they allow for reasonably accurate computation of Coulomb interactions at practicable computational costs. The reliance on a computational mesh means that there is a minimal length scale below which interaction cannot be resolved. The P3M algorithm provides a way to reintroduce interactions below the mesh size. This is achieved by computing the potential for close particles using direct summation techniques. In practice most codes rely directly on Coulomb's law, as given in eq (\ref{eq:summation_short}), where boundary conditions are not considered. 

In this paper we investigate the error when neglecting electrostatic boundary conditions in the computation of the potential at inter-particle distances smaller than the mesh size. 

Commonly, P3M is described using a splitting of the interaction, which is rooted in Ewald's summation. In our approach we use a mathematically equivalent splitting of the charge distribution through Gaussian charge screening. On the grounds of Low's Lagrangian (\ref{eq:lowlagrangian}) this allowed us to propose the Ansatz (\ref{eq:Lagrangian_Ansatz}), which lies at the core of this work. The discrete aspects of the computation were introduced by using macro-particles, grid interpolation functions and a leap-frog variational integrator respectively.\ 
Since the potential calculation of P3M involves many numerical steps the error is difficult to trace. For the purpose of quantifying the error we relied on a compact error measure (\ref{eq:error_measure}), based on the one used by \textcite{hockney_computer_1988}.

For the first time, a variational description of the P3M algorithm is presented, in the form of the discrete three-point Lagrangian given in eq (\ref{eq:full_p3m_lagrangian}). By applying the discrete Euler-Lagrange equation (\ref{eq:disc_euler_lagrange_eqn}), temporally and spatially discrete equations of motion (\ref{eq:eom_p3m_coord}, \ref{eq:eom_p3m_poiss_L}) and (\ref{eq:eom_p3m_poiss_S}) are derived.\ These results allow to obtain eqns (\ref{eq:phi_L}), (\ref{eq:phi_S}) for the potentials, as computed in the P3M scheme.\
The error made by neglecting the boundary conditions in the short range interaction is shown to be directly linked to the discretization error in the long range interaction. If the latter is small enough, the effect of boundary conditions in the short range interaction is suppressed, as long as particles are more than half an Ewald cut-off length away from a conductor (see fig \ref{fig:ShortError}). Usually, the Ewald cut-off is below or on the order of the mesh size, so this poses no loss in generality. 

In conclusion, it was proven that,

\begin{mdframed}[style=EndFrame]
\begin{align*}
    & G^L\ \text{minimizes discretization error eq (\ref{eq:error_measure})} \\ & \iff |G^S-G^F|=0\ \text{within the Ewald sphere}.
\end{align*} 
\end{mdframed}





\bibliography{Paper_refs.bib}

\end{document}